\documentclass[10pt, onecolumn]{extarticle}

\usepackage[bordercolor=white,backgroundcolor=gray!30,linecolor=black,colorinlistoftodos, textsize=10pt]{todonotes}

\usepackage{amssymb}
\usepackage{mathtools}
\usepackage{textcomp}
\usepackage{upgreek}
\usepackage{amsmath}
\usepackage[textstyle,amssymb]{SIunits}
\usepackage{graphicx}
\usepackage[normalem]{ulem}  % for underline allowing linebreaks
\usepackage{soul}
\usepackage[twocolumn,textwidth=183mm,columnsep=6mm,top=20mm,bottom=25mm]{geometry}
\usepackage{helvet}
\usepackage{titlesec}
\titleformat*{\section}{\bfseries\sffamily}
\titlespacing{\section}{0pt}{*4}{*0}
\titleformat{\subsection}[runin]{\normalfont\bfseries}{\thesubsection.}{3pt}{}
\usepackage{setspace}
\usepackage[breaklinks=true]{hyperref} %[breaklinks=true]
\usepackage[hang,flushmargin]{footmisc} 
\usepackage[font={footnotesize,sf},labelfont={sf,bf},labelsep=endash,justification=raggedright]{caption}

\usepackage{natbib}
\makeatletter
\def\blfootnote{\gdef\@thefnmark{}\@footnotetext}
\makeatother

\usepackage[labelfont=bf,justification=justified]{caption}

\begin{document}

\twocolumn[
\begin{@twocolumnfalse}
	%******************* title *******************
	{\LARGE \sf \textbf{Broadband THz wave generation and detection \\in organic crystal PNPA at MHz repetition rates}}
	\vspace{0.4cm}
	
	%******************* authors *******************
	{\sf\large \textbf {Lukasz A. Sterczewski$^{1,*}$, Jakub Mnich$^{1}$, and Jaroslaw Sotor$^{1}$}}
		\vspace{0.5cm}

		{\sf \textbf {\noindent $^1$Laser \& Fiber Electronics Group, Faculty of Electronics, Photonics and Microsystems, Wroclaw University \\of Science and Technology, Wyb. Wyspianskiego 27, 50-370 Wroclaw, Poland}
        }
		\vspace{0.5cm}
		
\normalsize
\end{@twocolumnfalse}]

%******************* abstract *******************

{\noindent \sf \small \textbf{\boldmath
Organic nonlinear optical (NLO) crystals have emerged as efficient room-temperature emitters of broadband THz radiation with high electric field strengths. Although initially confined to high-pulse-energy excitation in the millijoule range with kHz rates, recent efforts have focused on tailoring the physical properties of organic NLO materials for compatibility with popular MHz-rate telecommunication wavelength lasers emitting nanojoule energy pulses. This is motivated by the large potential of such crystals for more portable and user-safe spectroscopic systems, additionally complemented by their room-temperature field-sensitive THz detection capabilities. In this work, we demonstrate the MHz-rate operation of the organic crystal PNPA ((E)-4-((4-nitrobenzylidene)amino)-$N$-phenylaniline), which was recently discovered through data mining algorithms and reported to surpass the conversion efficiency of rivaling NLO crystals at 1~kHz repetition rate. Using a compact 200~mW Er:fiber laser producing 18~fs pulses at 50~MHz repetition rate, we demonstrate the THz field generation and detection capabilities of PNPA via performing gas phase spectroscopy in the 1.3--8.5~THz range. We obtain a dynamic range of 40~dB over 3.6~s and 70~dB over 2 hours. Our work extends the family of organic crystals compatible with telecommunication-wavelength excitation using nJ pulses for spectroscopy beyond 5~THz.}}

\blfootnote{\noindent$^*${lukasz.sterczewski@pwr.edu.pl}}

\vspace{0.5cm}

%____________________________________
\section{Introduction}
Organic NLO crystals have received considerable attention in the recent decade due to their broadband THz wave generation capabilities~\cite{jazbinsekOrganicCrystalsTHz2019}. They fill an important niche in the source spectral coverage by complementing the region above 5 THz, which is difficult to access using typical photoconductive switches owing to parasitic electrical parameters or insufficiently fast carrier dynamics~\cite{gobelTelecomTechnologyBased2013, turanWavelengthConversionPlasmoncoupled2021}, or using quantum cascade lasers (QCLs) due to photon-phonon coupling -- the \emph{Restrahlen} band effect~\cite{kohler2002terahertz, williams2007terahertz,shahiliContinuouswaveGaAsAlGaAs2024}. Although a plethora of different THz generation schemes exists, organic NLO crystals feature many practical advantages like compatibility with popular optical excitation wavelengths at moderate pump powers, ease of fabrication (growth), and operation at room temperature. 

Similar to inorganic crystals like gallium phosphide (GaP) or zinc telluride (ZnTe), which have traditionally been used for THz spectroscopy~\cite{fergusonMaterialsTerahertzScience2002}, organic NLO crystals offer broadband phase matching~\cite{jazbinsekOrganicCrystalsTHz2019}  reaching more than 10~THz (up to 30~THz with some dips) for crystals like DAST~\cite{heHighenergyUltrawidebandTunable2018}, DSTMS~\cite{sommaUltrabroadbandTerahertzPulses2015, pucUltraBroadbandHighDynamicRangeTHz2021}, OH-1~\cite{vicarioHighFieldBroadband2015} or BNA~\cite{zhaoEfficientBroadbandTerahertz2019}. The main difference, however, lies in their much higher NLO coefficients (up to 10$\times$) and larger pump wavelength flexibility. THz generation relies in these sources on optical rectification –- a second order nonlinear process, which requires one to match the group index at the pump wavelength (near-infrared) with the phase (refractive) index at THz frequencies (far-infrared). In most cases, sub-millimeter thicknesses are preferred  in organic NLO crystals due to the limited coherence length, above which the nonlinear conversion efficiency degrades. 

% Figure 1 - experimental setup
\begin{figure*}[!htp]
	\centering
	\includegraphics[width=.9\textwidth]{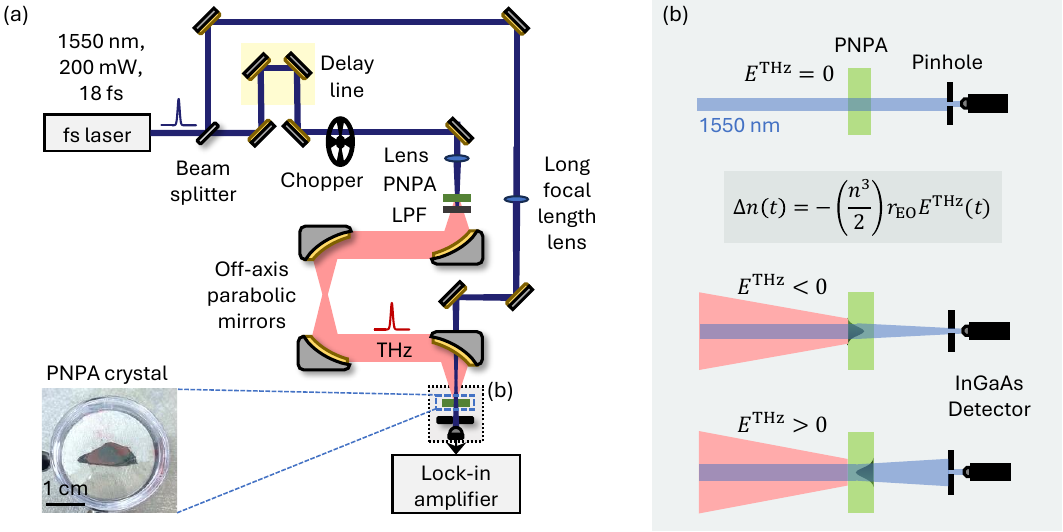}
	\caption{Experimental setup and principle of THz-induced lensing detection mechanism in PNPA a) Experimental setup constituting a THz time-domain spectrometer. (b) THz-induced lensing when no external THz field is present, when the field is positive (inducing a diverging lens), and when the field is negative (inducing a converging lens). In the formula, $r_\mathrm{EO}$ is the electro-optic coefficient (in pm/V), and $n$ is the refractive index at the probe wavelength (near-infrared).
	\label{fig:Schematic} }
\end{figure*}

What makes organic NLO crystals of even larger relevance is their bi-functionality – they can simultaneously serve as efficient field-sensitive detectors and emitters. Unlike in conventional electro-optic sampling systems, the detection process $\emph{does not}$ employ a field-induced birefringence. Instead, the Pockels effect is used, which is a linear change of the refractive index in the presence of an external THz $E$-field. It is responsible for the spatial modulation of the near-infrared probe beam incident on the crystal with a $\chi^{(2)}$ nonlinearity. This effect, known as THz-induced lensing (TIL), can be used for THz detection as proposed by Schneider et al.~\cite{schneider2004terahertz}. At high peak fields, the dynamically induced lens focuses or de-focuses the beam so that it can be detected by a small-aperture photodetector. Note that the Pockels effect considered here differs significantly from the Kerr effect, which is a third-order nonlinear process ($\chi^{(3)}$) proportional to field intensity ($|E(t)|^2$) that disregards the optical phase. 

From a practical perspective, a system using organic NLO crystals represents an elegant solution to broadband THz spectroscopy in amplitude and phase. One can use a lock-in detection scheme in tandem with sensitive near-infrared semiconductor photodetectors like InGaAs to probe very long wavelengths with high signal-to-noise ratios (SNR). Unfortunately, the dynamic range obtainable with organic NLO crystals is still far away from that of photoconductive antennas~\cite{vieweg2014terahertz, pucUltraBroadbandHighDynamicRangeTHz2021}, although for the latter obtainable at much lower frequencies (1~THz). 

To circumvent the limited conversion efficiency, phase matchability, and eliminate parasitic absorption dips in the emission spectrum, researchers have recently employed data mining algorithms to discover and develop new nonlinear materials for the THz range~\cite{valdivia2022data}. A successful result of this search is the new organic nonlinear crystal PNPA, which offers the highest intensity THz generation surpassing that of DAST or OH-1~\cite{raderCustomTerahertzWaveforms2022} or efficient second harmonic generation~\cite{hollandCharacterizationOrganicCrystals2023} and is commercially available. Its significant advantage is the native compatibility with telecommunication wavelengths (1550~nm), where convenient Er:fiber lasers can be used for optical excitation. 

Although PNPA holds a significant potential for broadband high-resolution THz spectroscopy, prior demonstrations have focused only the high-field THz generation properties with spectral coverage limited to 5~THz~\cite{raderNewStandardHighField2022}. Using mJ-energy 100~fs laser pulses with kHz repetition rates from an optical parametric amplifier (OPA), a nonlinear conversion efficiency exceeding 4\% has been demonstrated. Since OPA-based sources have rather large footprints and require many safety precautions, for widespread adoption of PNPA greatly desired is compatibility with compact femtosecond fiber lasers operating at telecommunication wavelengths with MHz rates.

In this work, we demonstrate not only the THz generation capabilities of PNPA in the low-power (200~mW average) 1550~nm excitation regime, but also uncover its previously unexplored field-sensitive detection capabilities. Despite the lowered conversion efficiency, pumping with nJ-level energy pulses with 50~MHz repetition rate ($\sim$5$\times$10$^4$ higher than reported to date) and 18~fs duration obtainable directly from the Er:fiber amplifier has enabled us to not only to double the coverage to above 10~THz, but also uncover quasi-discrete emission peaks at 18~THz and 23-24 THz. The spectral dynamic range obtainable over a second timescale reaches 40~dB, which improves to 70~dB upon coherent averaging over 2 hours. Our work extends the family of bi-functional organic NLO crystals like DAST or DSTMS~\cite{pucUltraBroadbandHighDynamicRangeTHz2021} suitable for broadband THz time-domain spectroscopy using compact telecommunication wavelength pump sources.

\section{Results and Discussion}
\subsection{Broadband THz generation}

To study the emission and previously unknown detection capabilities of the PNPA crystal using nJ energy pulses, we inserted two pieces of crystal (from Terahertz Innovations, USA, 0.8~mm thickness) into a home-made THz time-domain spectrometer (THz-TDS) to serve as a THz emitter and detector. For comparative purposes, we also used a DSTMS crystal ((4-$N$,$N$-dimethylamino-4’-$N$’-methyl-stilbazolium 2,4,6-trimethylbenzenesulfonate, from Swiss Terahertz, Switzerland) to detect THz waveforms produced by PNPA and vice versa. The experimental setup is shown in Fig.~\ref{fig:Schematic}a, where a femtosecond laser system with a 18~fs pulse duration, 200~mW of average power, and 50~MHz repetition rate is collimated to a beam waist of 2~mm (1/e$^2$) and next split by a thin 90/10 beam splitter into pump and probe beams. Figure~\ref{fig:Schematic}b plots the concept of TIL detection for convenience. 

\begin{figure*}[!tb]
	\centering
	\includegraphics[width=.95\textwidth]{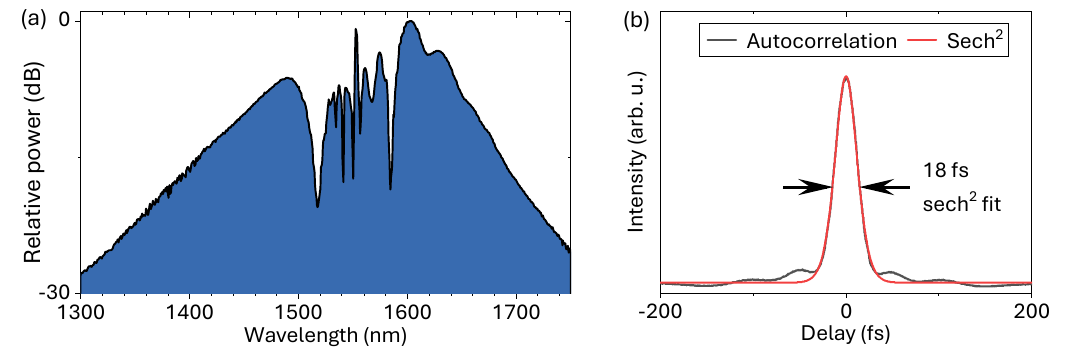}
	\caption{Er:fiber laser system used as a near-infrared pump for the generation and detection of THz radiation using PNPA. (a) Optical spectrum after the Er amplifier with characteristic features of self-phase modulation responsible for spectral broadening. (b) Intensity autocorrelation along with a sech$^2$ fit indicating a pulse duration of 18~fs.}
	\label{fig:Pump}
\end{figure*}

To ensure the broadest spectral coverage in the THz, it is essential to keep the near-infrared excitation pulse as short as possible. Fig.~\ref{fig:Pump}a shows the measured broadband emission spectrum owing its shape to self-phase modulation (SPM) occurring in the amplifier. A result of in-fiber post-compression following the SPM process is the autocorrelation shown in Fig.~\ref{fig:Pump}, which is clean and almost free of side lobes signifying most of the power concentrated in the main pulse. To excite the PNPA emitter, the pump arm is mechanically modulated by an optical chopper at 5~kHz and delayed by a precise mechanical stage moving a hollow roof prism mirror relative to the probe arm. To focus the beam onto the crystal, we use an $f$=75~mm achromatic doublet lens (glass materials with opposite-sign group velocity dispersions), which produces a 2$w_\mathrm{opt}$=75~$\upmu$m wide focal spot with a 5.6~mm depth of field. The latter allows us to move the crystal relative to the first off-axis parabolic mirror without significantly changing the pump fluence. Upon illuminating the PNPA crystal, the pump beam is blocked by a 5~mm thick germanium window serving as a long-pass filter (LPF). The THz pulse passes through the filter to be next imaged onto the detector crystal using 4 gold-coated off-axis parabolic mirrors ($f$=50.8~mm). The last mirror preceding the detection stage has a hole allowing to spatially overlap the THz pulse with the probe.

\subsection{Broadband THz detection}
For the detection, we weakly focus the probe beam to $\sim$200~$\upmu$m by a thin, $f$=200~mm plano-convex lens, which provides a quasi-collimated beam with a $\sim$40~mm depth of field. Empirically, 50~mm away from the crystal we found the beam diameter to increase its diameter from the focus by $\sqrt{2}$ to a beam waist of 300~$\upmu$m, which defines our empirical Raleigh range. We placed there an unbiased InGaAs detector with a 100~$\upmu$m pinhole to measure only the central part of the beam intensity, as emphasized in the seminal work in Ref.~\cite{schneider2004terahertz}. Such a simple detection geometry avoids the need for sophisticated beam telescopes to lower the beam diameter and has proven to provide quality THz spectroscopy data in agreement the HITRAN~\cite{gordon2022hitran2020} spectroscopic database (discussed later in the text). For convenience, we provide a conceptual drawing of TIL-based detection scheme in Fig.~\ref{fig:Schematic}b. Since the AC component of the detected photocurrent lies in the nA range, we used a low-noise transimpedance amplifier (DHPCA-100) to convert the detected current into voltage detected in a homodyne configuration using a lock-in amplifier. 

\begin{figure*}[!tb]
	\centering
	\includegraphics[width=.95\textwidth]{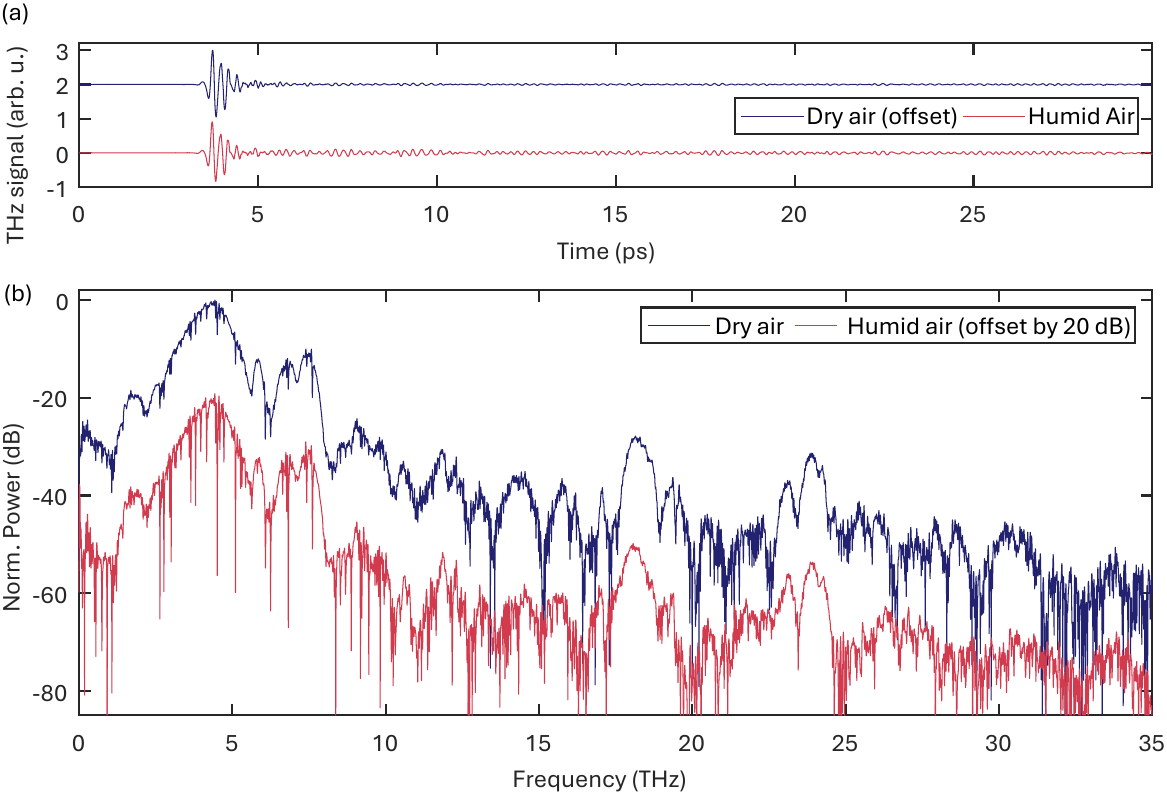}
	\caption{THz spectroscopy experiment wherein the measurement chamber was purged with argon. (a) Time-domain waveform normalized to the water-vapor free THz signal. (b) Corresponding THz power spectra offset by 20~dB for clarity. The propagation environment was technical argon and laboratory air (15\% relative humidity), 0.43\% H$_2$O by volume.}
	\label{fig:PurgingExperiment}
\end{figure*}

\subsection{THz spectroscopy of humid air}
To test the spectroscopic capabilities of PNPA, we enclosed the time-domain spectrometer in a hermetic chamber equipped with a humidity sensor. First, we recorded the THz spectrum without purging the propagation environment followed by its intensive purging with argon. When water vapor lines almost completely disappeared from the THz spectrum, suggesting a relative humidity lower than 2\%, we acquired a second THz spectrum. Results of the experiment are plotted in Fig.~\ref{fig:PurgingExperiment}. The acquisition delay range was 30~ps, which corresponds to a theoretical optical resolution of 33.3~GHz ($\sim$1.11~cm$^{-1}$). To obtain the spectrum, 10 step scans with an integration time of 30~ms per step were averaged.

It is evident from Fig.~\ref{fig:PurgingExperiment}a that removing water vapor from the measurement chamber through purging greatly suppresses THz field oscillations after the main pulse. It is expected since sharp absorption dips in the frequency spectrum must correspond to long-lasting sine-wave-like oscillations in the time domain. It should be noted that it is only the spectral region up to 10~THz (30~$\upmu$m of wavelength) that is attenuated by H$_2$0. Weak emission peaks around 18~THz and higher frequencies do not change in the presence of argon instead of laboratory air.

\begin{figure*}[!tb]
	\centering
	\includegraphics[width=1\textwidth]{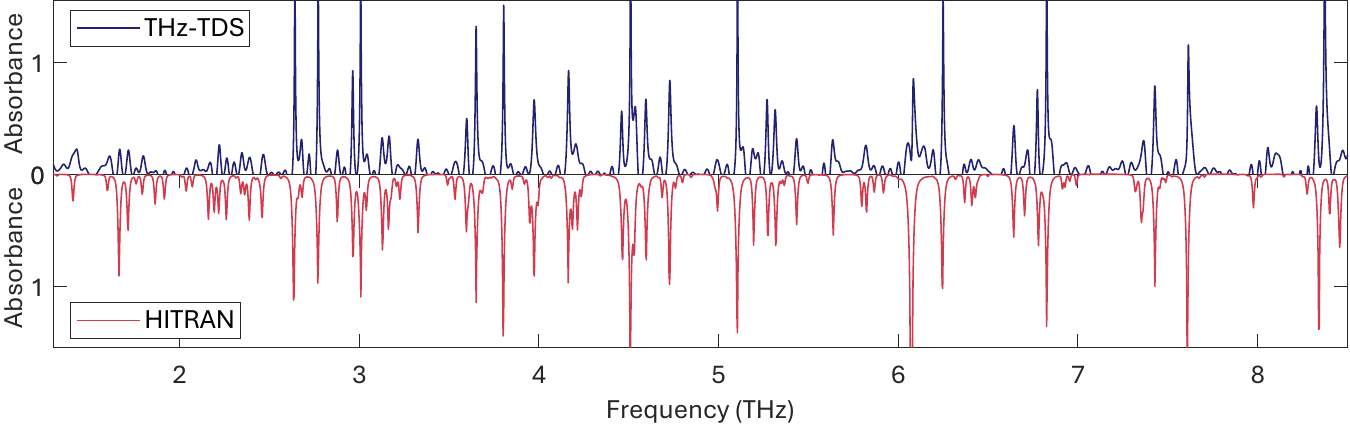}
	\caption{Measured absorbance spectrum of humid air (15\% relative humidity) in the measurement chamber using THz-TDS plotted with a HITRAN2020 model~\cite{gordon2022hitran2020}.}
	\label{fig:H20_spectroscopy}
\end{figure*}

\begin{figure*}[!htp]
	\centering
	\includegraphics[width=.9\textwidth]{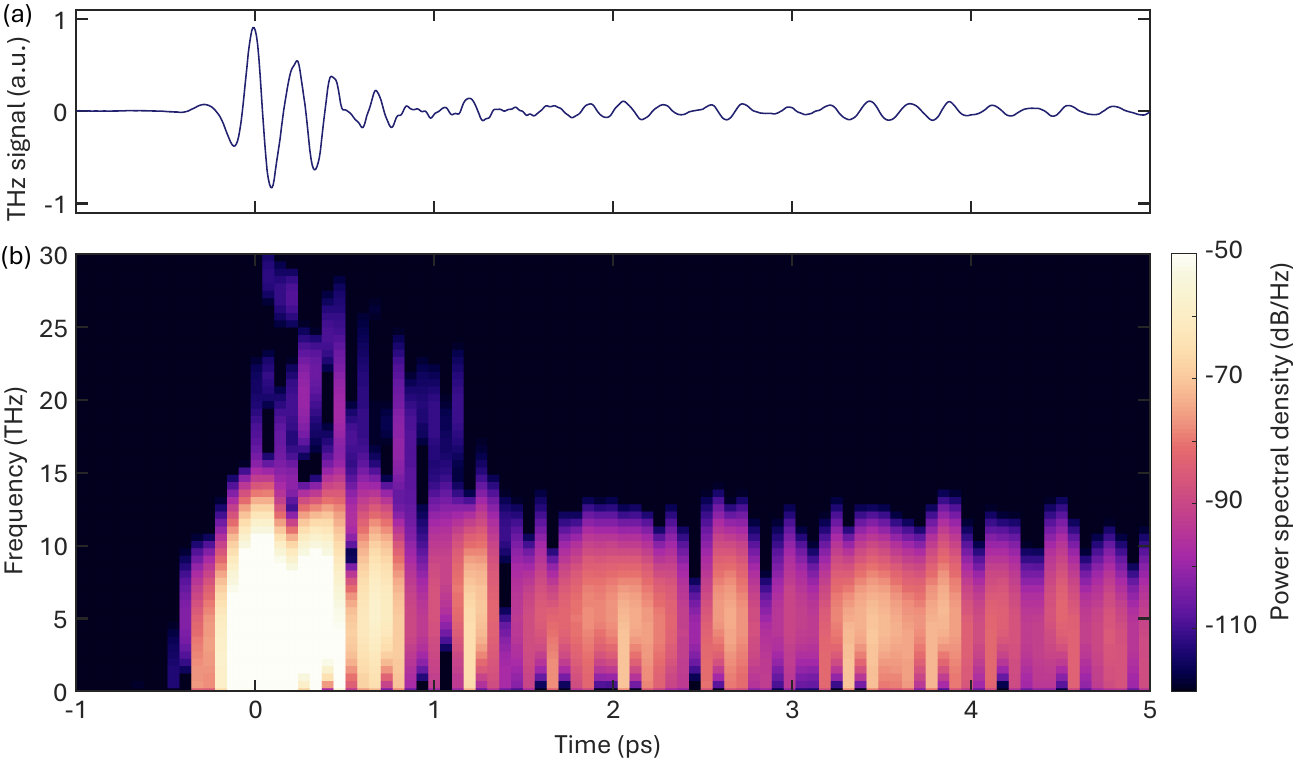}
	\caption{Spectrogram of the PNPA emission waveform after propagation in ambient air. (a) Sampled $E$-field. (b) Corresponding spectrogram displaying a chirped nature of the signal -- lower frequencies appear first with a gradual broadening towards 20--30~THz (15--10~$\upmu$m of wavelength). After 1~ps, only oscillations below 10~THz persist, which arise due to the rotational motion of water vapor molecules along the propagation path of the THz pulse.}
	\label{fig:spectrogram}
\end{figure*}

By subtracting base-10 logarithmic spectra from Fig.~\ref{fig:PurgingExperiment}b, we generated the absorbance spectrum plotted in Fig.~\ref{fig:H20_spectroscopy}. Good agreement with the HITRAN database is obtained (USA model, mean latitude, winter. 0.3~cm$^{-1}$ apodized resolution, Voigt profile with a triangular instrumental function). This unequivocally proves the PNPA's ability to serve as a material for broadband THz spectroscopy with moderate excitation powers obtainable from Er:fiber laser systems. Minor discrepancies between the measured spectrum and the model relate to the finite dynamic range of the instrument varying across the spectrum (the highest at $\sim$5~THz), and insufficiently precise sample triggering mechanism. Here we relied on the built-in encoder of the delay line, but further extension of the spectrum to the mid-infrared range may require interferometric, laser-based sample triggering.

\subsection{Spectro-temporal characteristics}

Access to the THz $E$-field (its scaled representation convolved with the detector- and instrumental response function) allows us to study the spectro-temporal characteristics of the generated signal. An analysis performed using a short-time Fourier transform is shown in Fig.~\ref{fig:spectrogram}, where the top panel plots the THz waveform (Fig.~\ref{fig:spectrogram}a), while the bottom includes is its spectrogram representation Fig.~\ref{fig:spectrogram}b). As evident from the figure, the waveform is chirped with lower frequencies reaching the detector first, followed by a gradually broadened spectral response with frequency components above 20~THz. After $\sim$1~ps they disappear from the spectrum and only oscillations from water vapor molecules persist. The intensive part of the pulse is approximately three optical cycles long and lasts less than 1~ps. The origin of the waveform chirp can be attributed to dispersive properties of the crystal (also for the probe) and of the Ge filter with the former being more likely to dominate. The waveform's temporal structure may also originate from the molecular dynamics of the crystal, where the oscillatory part following the main peak stems from collective vibrations of the molecular system~\cite{sommaUltrabroadbandTerahertzPulses2015}. They spectrally interfere with the broadband single-cycle THz pulse resulting directly from pump excitation and give rise to a modulated spectrum with prominent peaks and regions with a lack of THz power due to destructive interference~\cite{singh2020up}.

It is important to note that the peak emission frequency is also dependent on the beam waist ($w_\mathrm{opt}$) stimulating the emission crystal~\cite{jazbinsekOrganicCrystalsTHz2019}. For tightly-focused beams, lower generation frequencies suffer more from diffraction losses. For instance, for a focal spot of 75~$\upmu$m, at 2~THz we anticipate 20~dB of THz power losses, which drops to 15 dB at 5~THz and 11~dB at 10~THz. Doubling the beam diameter to 150~$\upmu$m lowers these values to 12~dB, 7~dB, and 4~dB, respectively. On the other hand, the peak intensity of the pump drops quadratically with the beam diameter $w_\mathrm{opt}$. Since the THz field is proportional to $E_\mathrm{p}/(\frac{1}{2}\mathrm{\pi}w_\mathrm{opt}^2)$, where $E_\mathrm{p}$ is the pulse energy, increasing the beam diameter $k$ times lowers the obtainable THz power by a factor or $k^4$. By simply doubling $k$ while keeping other pump parameters unchanged, the THz power lowers by 12~dB. This explains why in the mJ pulse energy regime, THz emission spectra show generally more low-frequency content~\cite{raderCustomTerahertzWaveforms2022, raderNewStandardHighField2022} than with nJ pulse excitation reported here and in Ref.~\cite{pucUltraBroadbandHighDynamicRangeTHz2021, puc2023broadband}. It relates to a trade-off between the beam size limited by the damage threshold (which should be as small as possible for high conversion efficiency) and diffraction losses affecting predominantly lower frequencies for tightly focused beams.

\begin{figure*}[!tb]
	\centering
	\includegraphics[width=1\textwidth]{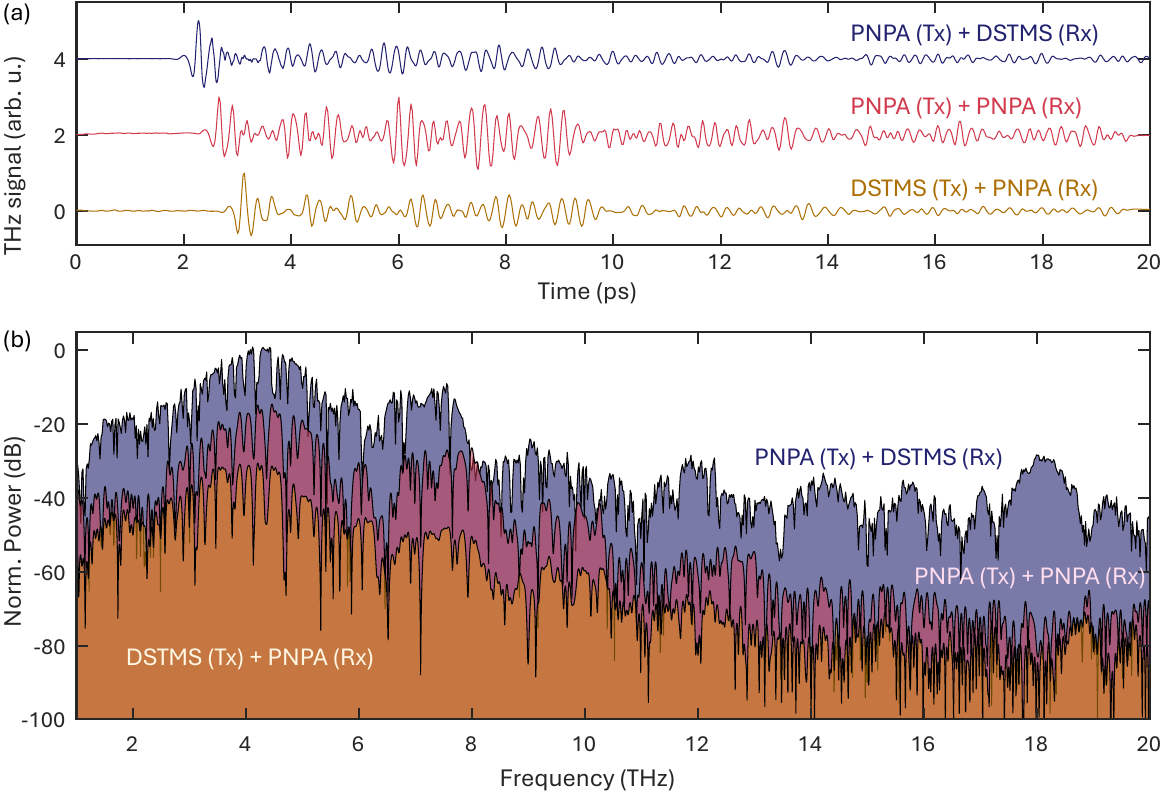}
	\caption{THz spectroscopy using different generation (Tx) and detection crystal (Rx) pairs. (a) Time-domain THz signal. (b) Power spectra plotted with a 20~dB offset. Although DSTMS offers the broadest detection spectrum, in the relevant spectral region (2--8~THz), PNPA can serve equally well for this purpose.}
	\label{fig:DetectionUsing}
\end{figure*}

\subsection{Detection capabilities}
Until now, we used a well-characterized, broadband detection crystal -- DSTMS to study the broadband emission properties of PNPA with sub-20-fs pulses. This is because unlike for electrooptic sampling, retrieving the $E$-field strength using TIL is cumbersome as the measured quantity is only the relative change of intensity ($\Delta I/I$). Therefore, a reasonable approach to assess the performance of organic NLO detectors is comparing new materials with well-documented molecular standards or other inorganic crystals~\cite{schneider2004terahertz}. 

Following this reasoning, first we performed a series of measurements with PNPA and DSTMS interchangeably serving as a source and detector to finally compare PNPA with PNPA. Results of time-domain measurements are plotted in Fig.~\ref{fig:DetectionUsing}a along with corresponding power spectra in Fig.~\ref{fig:DetectionUsing}b. 

Since the time-domain waveforms are acquired over a relatively long delay range of 20~ps, water vapor absorption features are prominent. Even though the spectra are plotted with a 20~dB offset, emission from the PNPA crystal detected by DSTMS offers the broadest spectral response with frequency components filling the entire spectral region of interest (20~THz). On the other hand, within the region of the highest spectral power (below 8~THz), minimal differences between the spectra exists. This proves the PNPA's suitability for broadband THz spectroscopy.

\begin{figure*}[!tb]
	\centering
	\includegraphics[width=0.75\textwidth]{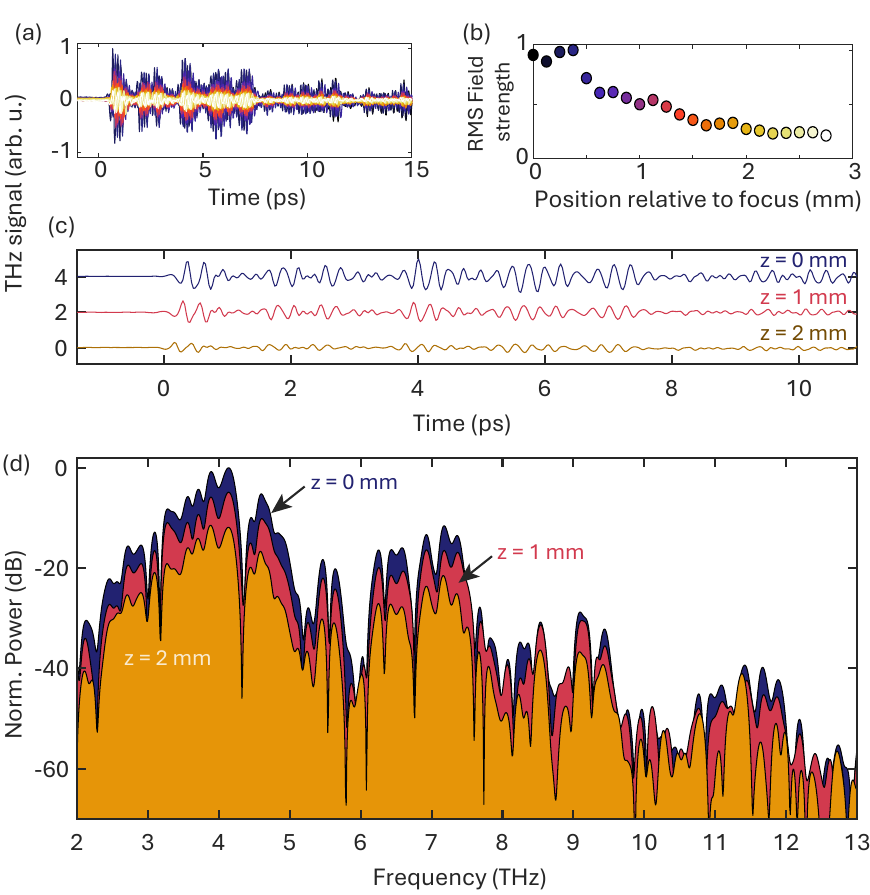}
	\caption{Sensitivity of the detection system to axial shift when the PNPA detection crystal was moved in 0.125~mm steps from 0 to 2.75~mm from the focus of the off-axis parabolic mirror. (a) Stack of time-domain traces. (b) RMS field strength calculated from (a). (c) 3 traces at arbitrary positions: 0 mm (in focus), offset by 1~mm, and 2~mm. (d) Corresponding power spectra.}
	\label{fig:Focusing}
\end{figure*}

\subsection{Conversion efficiency}
To assess the performance more quantitatively, we measured the average THz power using a broadband LiTaO$_3$ pyroelectric detector with a HDPE window (Laser Components, Germany). From DSTMS we measure 430~nW (after the Ge window), which corresponds to a power conversion efficiency of $\eta_\mathrm{D}$=2.4$\cdot$10$^{-6}$. For PNPA we measure an average power of 170~nW, which yields $\eta_\mathrm{P}$=9.4$\cdot$10$^{-7}$. This discrepancy, or at least the superiority of DSTMS over PNPA in terms of THz power conversion efficiency, can potentially be explained by the lack of broadband phase matching in the case of PNPA with frequency components extending over 30~THz rather than 10~THz~\cite{jazbinsekOrganicCrystalsTHz2019}. Also, prior data on PNPA as an emitter focused on the lower-frequency THz region limited to 5~THz with most of the power concentrated at 1.5--2.5~THz. 

Pumping with mJ-energy pulses often leads to percent-level nonlinear conversion efficiency, while for telecom-wavelength Er:fiber systems $\eta$ on the order of 10$^{-6}$ is naturally expected. We can explain it using the linear dependence of the THz field with respect to pulse energy, which translates into a quadratic dependence for the THz power. Accounting for the large-area illumination ($w_\mathrm{opt}$=3.8~mm, $k$=50 larger than in our setup), and 0.6 mJ pulse energy (vs 3.6~nJ here, 1.7$\cdot$10$^5$ higher) we anticipate $\sim$10$^5$ THz more power than in our setup, which is close to the mW average THz power range, when percentage level conversion efficiency was obtained. We want to emphasize that it is only a coarse approximation ignoring saturation or thermal effects, yet it explains the obtained performance metrics.

\subsection{Detection sensitivity}
To characterize the detection sensitivity we measure the relative intensity (photocurrent) change induced by the THz field produced by DSTMS and PNPA. With average measured photocurrents (DC) lying in the $\upmu$A range, we measure peak THz-induced photocurrent changes in the nA range. Specifically, $\Delta I/I$=6$\cdot$10$^{-4}$ for both DSTMS and PNPA serving as detection crystals. When PNPA crystal was used for the generation and detection (with 2.5$\times$ less average power), $\Delta I/I$=3.4$\cdot$10$^{-4}$. Considering the experimental uncertainty due to alignment and nearly identical refractive indices of DSTMS and PNPA~\cite{hollandCharacterizationOrganicCrystals2023}, we conclude similar detection performance of PNPA to that of DSTMS. A more accurate assessment of the detection capabilities (including the electro-optic coefficient $r_\mathrm{EO}$) will require refractometry of PNPA in tandem with calibration of the $E$-field strength, as required by the Pockels effect formula from Fig.~\ref{fig:Schematic}b. From the current measurements, the most obvious observation is the lack of ultra-broadband detection ($>$10~THz) capabilities of PNPA at telecommuncation wavelengths. However, phase matching may be present closer to the mid-infrared pump region, where very few organic NLO materials have been proven to operate~\cite{gollner2021highly}.

Another practical aspect of the TIL scheme is the sensitivity to axial position of the detection crystal with respect to the THz-focusing off-axis parabolic mirror. In steps of 0.125~$\upmu$m, we moved the position of PNPA relative to the point of maximum detected THz amplitude over 2.75~mm. Recorded THz signals (Fig.~\ref{fig:Focusing}a,c) along with the normalized RMS field strength (Fig.~\ref{fig:Focusing}b) and power spectra (Fig.~\ref{fig:Focusing}d) indicate a rapid drop of the detected field strength when the crystal was moved by more than 0.5~mm away from the focal spot. On the other hand, even at large distances (2~mm), the THz waveform was still detectable, albeit with $\sim$14~dB loss (0.2 of the maximum field strength). It is important to note that almost all THz frequencies attenuate by the same amount with little-to-no visual difference between them, which is not expected for broadband polychromatic light detected using the TIL technique. 

The behavior observed in Fig.~\ref{fig:Focusing} can be explained considering the spot size of the THz beam $w_0=w_\mathrm{opt}$/$\sqrt{2}$ (the $\sqrt{2}$ factor is because the THz field is directly proportional to the pump intensity $I$, hence the THz power is proportional to $I^2$~\cite{pucUltraBroadbandHighDynamicRangeTHz2021}). Empirically, we measured the THz spot size using a pyroelectric detector with a calibrated pinhole, which in our case amounted to $\sim$150~$\upmu$m ($\pm$50~$\upmu$m). Such a beam has a diameter for the $E$-field of $\sim$200~$\upmu$m, which doubles at 0.5~mm away from the focus, and increases its size 7 times 2~mm away considering Gaussian beam propagation. This is in good agreement with the data shown in Fig.~\ref{fig:Focusing}b, but also proves that our THz imaging system (using 4 off-axis parabolic mirrors) is not far from diffraction-limited performance if one assumes a peak emission wavelength of 70~$\upmu$m and a collimated THz beam diameter of $\sim$35~mm with a theoretical spot size of 130~$\upmu$m.

\subsection{Dynamic range capabilities}
\begin{figure}[!tb]
	\centering
	\includegraphics[width=1\columnwidth]{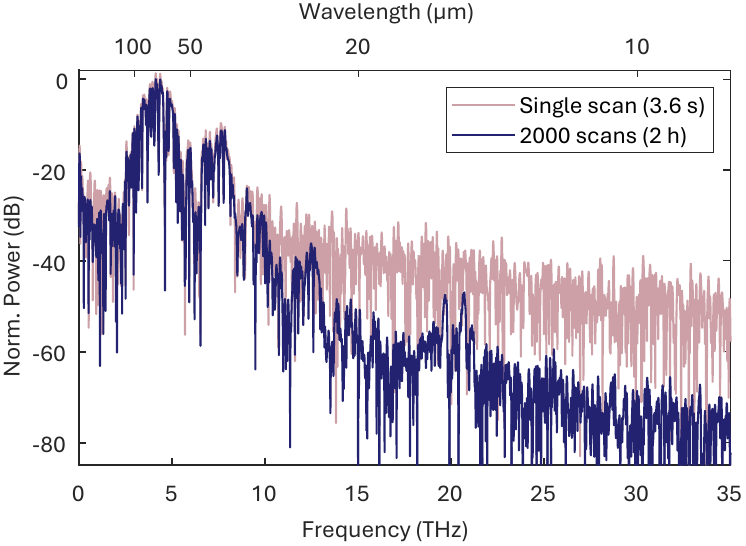}
	\caption{Dynamic range capabilities of the PNPA crystal serving as a source and detector.}
	\label{fig:Averaging}
\end{figure}

Finally, we characterized the dynamic range capabilities of the THz-TDS system employing PNPA crystals. 20~ps long traces were acquired in rapid acquisition mode with a lock-in integration time of 1~ms. Each trace consisted of 1600 points taken every 3.6 seconds. Via cross-correlation-based alignment of THz traces (to correct for the random jitter of the acquisition trigger event), we obtained 2000 signals averaged in the time domain. Fig.~\ref{fig:Averaging} shows the dynamic range capabilities, where a single 3.6~s-long scan provides $\sim$40~dB of dynamic range, which improves to $\sim$70~dB upon integration. We want to emphasize, however, that scans on a minute timescale already provide dozens of dB of dynamic range improvement from 40 to almost 60~dB, which should already be sufficient for many spectroscopic investigations. 

\section{Conclusion}
In summary, we have demonstrated the nJ pulse energy compatibility of the organic NLO crystal PNPA with an Er:fiber laser system to generate and detect field-resolved THz waveforms. Using 18-fs long pulses with 200~mW of average power centered at $\sim$1.56~$\upmu$m of wavelength, we have obtained 170~nW of THz power spread over more than 20~THz with the most intensive features in the 2--8~THz range. The emitted waveform exhibits slight chirp potentially attributable to optical dispersion and molecular system dynamics. We have also shown the THz-induced lensing capabilities of PNPA similar to DSTMS rendering a peak spectral dynamic range of 40~dB over seconds and 70~dB over 2~h of averaging. Although PNPA lacks the ultra-broadband emission characteristics of rivaling organic NLO crystals when pumped by 1550~nm sources, it is likely to be preferred for sub-10-THz frequency applications or to complement dips in the emission spectra from crystals like DSTMS or DAST via THz waveform synthesis~\cite{raderCustomTerahertzWaveforms2022}. 

We also want to note that ultra-broadband (70~THz coverage) Ge-based photoconductive antennas~\cite{singh2018gapless} have recently been made compatible with Er:fiber laser technology~\cite{singh2020up}. As a competing platform, they show a lot of promise for gap-less THz emission. Unfortunately, when pumped at 1550~nm (instead of the optimal 1100~nm), they lack power between 3--10~THz. This is where PNPA can potentially serve to fill the spectral gap.

\section{Experimental Section}

\vspace{0.2cm}
\footnotesize
\noindent\textbf{Calculation of Fourier spectra}: Each time-domain waveform was windowed using a Tukey window with a shape parameter of 0.1 after mean value removal. 32-times zero padding has been used to obtain interpolated, smooth power spectra using the fast Fourier transform (fft) algorithm. 

\footnotesize
\vspace{0.2cm}

%************** Acknowledgement **************
%\vspace{-0.2cm}
\section*{Acknowledgements}
\footnotesize
\noindent
L. A. Sterczewski and J. Mnich acknowledge funding from the European Union (ERC Starting Grant, TeraERC, 101117433). Views and opinions expressed are however those of the authors only and do not necessarily reflect those of the European Union or the European Research Council Executive Agency. Neither the European Union nor the granting authority can be held responsible for them. This work is supported by the use of the National Laboratory for Photonics and Quantum Technologies (NPLQT) infrastructure, which is financed by the European Funds under the Smart Growth Operational Programme. The authors would like to thank the Swiss Terahertz team (Z{\"u}rich, Switzerland) for valuable technical discussions. Prof. Mojca Jazbinsek (Z{\"u}rich University of Applied Sciences, Switzerland) is acknowledged for help with the THz-induced lensing detection technique.  

%\vspace{-0.2cm}
\section*{Author contributions}
\footnotesize
\noindent L.A.S. conceived the idea. J.S. fabricated the femtosecond laser. L.A.S. carried out the optical and electrical measurements with help from J.M., and analyzed the data. L.A.S wrote the manuscript with input from J.M. and J.S.. L.A.S. acquired funding and coordinated the project.

%\vspace{-0.2cm}
\section*{Conflict of interest}
\footnotesize
\noindent The authors declare no conflict of interest.

\section*{Data availability statement}
\footnotesize
\noindent Data are available in a public, open access repository. The data that support the findings of this study are openly available in figshare: \url{https://dx.doi.org/10.6084/m9.figshare.26304973}.

%\section*{Supplementary information}
%Supporting Information is available from the Wiley Online Library or from the author.

% References
\medskip

\bibliographystyle{naturemag}
\bibliography{main}

\begin{thebibliography}{10}
\expandafter\ifx\csname url\endcsname\relax
  \def\url#1{\texttt{#1}}\fi
\expandafter\ifx\csname urlprefix\endcsname\relax\def\urlprefix{URL }\fi
\providecommand{\bibinfo}[2]{#2}
\providecommand{\eprint}[2][]{\url{#2}}

\bibitem{jazbinsekOrganicCrystalsTHz2019}
\bibinfo{author}{Jazbinsek, M.}, \bibinfo{author}{Puc, U.}, \bibinfo{author}{Abina, A.} \& \bibinfo{author}{Zidansek, A.}
\newblock \bibinfo{title}{Organic {{Crystals}} for {{THz Photonics}}}.
\newblock \emph{\bibinfo{journal}{Applied Sciences}} \textbf{\bibinfo{volume}{9}}, \bibinfo{pages}{882} (\bibinfo{year}{2019}).

\bibitem{gobelTelecomTechnologyBased2013}
\bibinfo{author}{G{\"o}bel, T.} \emph{et~al.}
\newblock \bibinfo{title}{Telecom technology based continuous wave terahertz photomixing system with 105 decibel signal-to-noise ratio and 35 terahertz bandwidth}.
\newblock \emph{\bibinfo{journal}{Optics Letters}} \textbf{\bibinfo{volume}{38}}, \bibinfo{pages}{4197} (\bibinfo{year}{2013}).

\bibitem{turanWavelengthConversionPlasmoncoupled2021}
\bibinfo{author}{Turan, D.} \emph{et~al.}
\newblock \bibinfo{title}{Wavelength conversion through plasmon-coupled surface states}.
\newblock \emph{\bibinfo{journal}{Nature Communications}} \textbf{\bibinfo{volume}{12}}, \bibinfo{pages}{4641} (\bibinfo{year}{2021}).

\bibitem{kohler2002terahertz}
\bibinfo{author}{K{\"o}hler, R.} \emph{et~al.}
\newblock \bibinfo{title}{Terahertz semiconductor-heterostructure laser}.
\newblock \emph{\bibinfo{journal}{Nature}} \textbf{\bibinfo{volume}{417}}, \bibinfo{pages}{156--159} (\bibinfo{year}{2002}).

\bibitem{williams2007terahertz}
\bibinfo{author}{Williams, B.~S.}
\newblock \bibinfo{title}{Terahertz quantum-cascade lasers}.
\newblock \emph{\bibinfo{journal}{Nature Photonics}} \textbf{\bibinfo{volume}{1}}, \bibinfo{pages}{517--525} (\bibinfo{year}{2007}).

\bibitem{shahiliContinuouswaveGaAsAlGaAs2024}
\bibinfo{author}{Shahili, M.} \emph{et~al.}
\newblock \bibinfo{title}{Continuous-wave {{GaAs}}/{{AlGaAs}} quantum cascade laser at 5.7 {{THz}}}.
\newblock \emph{\bibinfo{journal}{Nanophotonics}} \textbf{\bibinfo{volume}{13}}, \bibinfo{pages}{1735--1743} (\bibinfo{year}{2024}).

\bibitem{fergusonMaterialsTerahertzScience2002}
\bibinfo{author}{Ferguson, B.} \& \bibinfo{author}{Zhang, X.-C.}
\newblock \bibinfo{title}{Materials for terahertz science and technology}.
\newblock \emph{\bibinfo{journal}{Nature Materials}} \textbf{\bibinfo{volume}{1}}, \bibinfo{pages}{26--33} (\bibinfo{year}{2002}).

\bibitem{heHighenergyUltrawidebandTunable2018}
\bibinfo{author}{He, Y.} \emph{et~al.}
\newblock \bibinfo{title}{High-energy and ultra-wideband tunable terahertz source with {{DAST}} crystal via difference frequency generation}.
\newblock \emph{\bibinfo{journal}{Applied Physics B}} \textbf{\bibinfo{volume}{124}}, \bibinfo{pages}{16} (\bibinfo{year}{2018}).

\bibitem{sommaUltrabroadbandTerahertzPulses2015}
\bibinfo{author}{Somma, C.} \emph{et~al.}
\newblock \bibinfo{title}{Ultra-broadband terahertz pulses generated in the organic crystal {{DSTMS}}}.
\newblock \emph{\bibinfo{journal}{Optics Letters}} \textbf{\bibinfo{volume}{40}}, \bibinfo{pages}{3404} (\bibinfo{year}{2015}).

\bibitem{pucUltraBroadbandHighDynamicRangeTHz2021}
\bibinfo{author}{Puc, U.}, \bibinfo{author}{Bach, T.}, \bibinfo{author}{G{\"u}nter, P.}, \bibinfo{author}{Zgonik, M.} \& \bibinfo{author}{Jazbinsek, M.}
\newblock \bibinfo{title}{Ultra-{{Broadband}} and {{High-Dynamic-Range THz Time-Domain Spectroscopy System Based}} on {{Organic Crystal Emitter}} and {{Detector}} in {{Transmission}} and {{Reflection Geometry}}}.
\newblock \emph{\bibinfo{journal}{Advanced Photonics Research}} \textbf{\bibinfo{volume}{2}}, \bibinfo{pages}{2000098} (\bibinfo{year}{2021}).

\bibitem{vicarioHighFieldBroadband2015}
\bibinfo{author}{Vicario, C.}, \bibinfo{author}{Ruchert, C.} \& \bibinfo{author}{Hauri, C.}
\newblock \bibinfo{title}{High field broadband {{THz}} generation in organic materials}.
\newblock \emph{\bibinfo{journal}{Journal of Modern Optics}} \textbf{\bibinfo{volume}{62}}, \bibinfo{pages}{1480--1485} (\bibinfo{year}{2015}).

\bibitem{zhaoEfficientBroadbandTerahertz2019}
\bibinfo{author}{Zhao, H.} \emph{et~al.}
\newblock \bibinfo{title}{Efficient broadband terahertz generation from organic crystal {{BNA}} using near infrared pump}.
\newblock \emph{\bibinfo{journal}{Applied Physics Letters}} \bibinfo{pages}{5} (\bibinfo{year}{2019}).

\bibitem{schneider2004terahertz}
\bibinfo{author}{Schneider, A.}, \bibinfo{author}{Biaggio, I.} \& \bibinfo{author}{G{\"u}nter, P.}
\newblock \bibinfo{title}{Terahertz-induced lensing and its use for the detection of terahertz pulses in a birefringent crystal}.
\newblock \emph{\bibinfo{journal}{Applied Physics Letters}} \textbf{\bibinfo{volume}{84}}, \bibinfo{pages}{2229--2231} (\bibinfo{year}{2004}).

\bibitem{vieweg2014terahertz}
\bibinfo{author}{Vieweg, N.} \emph{et~al.}
\newblock \bibinfo{title}{Terahertz-time domain spectrometer with 90 {{dB}} peak dynamic range}.
\newblock \emph{\bibinfo{journal}{Journal of Infrared, Millimeter, and Terahertz Waves}} \textbf{\bibinfo{volume}{35}}, \bibinfo{pages}{823--832} (\bibinfo{year}{2014}).

\bibitem{valdivia2022data}
\bibinfo{author}{{Valdivia-Berroeta}, G.~A.} \emph{et~al.}
\newblock \bibinfo{title}{Data mining for terahertz generation crystals}.
\newblock \emph{\bibinfo{journal}{Advanced Materials}} \textbf{\bibinfo{volume}{34}}, \bibinfo{pages}{2107900} (\bibinfo{year}{2022}).

\bibitem{raderCustomTerahertzWaveforms2022}
\bibinfo{author}{Rader, C.} \emph{et~al.}
\newblock \bibinfo{title}{Custom terahertz waveforms using complementary organic nonlinear optical crystals}.
\newblock \emph{\bibinfo{journal}{Optics Letters}} \textbf{\bibinfo{volume}{47}}, \bibinfo{pages}{5985} (\bibinfo{year}{2022}).

\bibitem{hollandCharacterizationOrganicCrystals2023}
\bibinfo{author}{Holland, K.~M.} \emph{et~al.}
\newblock \bibinfo{title}{Characterization of organic crystals for second-harmonic generation}.
\newblock \emph{\bibinfo{journal}{Optics Letters}} \textbf{\bibinfo{volume}{48}}, \bibinfo{pages}{5855--5858} (\bibinfo{year}{2023}).

\bibitem{raderNewStandardHighField2022}
\bibinfo{author}{Rader, C.} \emph{et~al.}
\newblock \bibinfo{title}{A {{New Standard}} in {{High-Field Terahertz Generation}}: The {{Organic Nonlinear Optical Crystal PNPA}}}.
\newblock \emph{\bibinfo{journal}{ACS Photonics}} \textbf{\bibinfo{volume}{9}}, \bibinfo{pages}{3720--3726} (\bibinfo{year}{2022}).

\bibitem{gordon2022hitran2020}
\bibinfo{author}{Gordon, I.~E.} \emph{et~al.}
\newblock \bibinfo{title}{The {HITRAN2020} molecular spectroscopic database}.
\newblock \emph{\bibinfo{journal}{Journal of Quantitative Spectroscopy and Radiative Transfer}} \textbf{\bibinfo{volume}{277}}, \bibinfo{pages}{107949} (\bibinfo{year}{2022}).

\bibitem{singh2020up}
\bibinfo{author}{Singh, A.} \emph{et~al.}
\newblock \bibinfo{title}{Up to 70 {THz} bandwidth from an implanted ge photoconductive antenna excited by a femtosecond er: fibre laser}.
\newblock \emph{\bibinfo{journal}{Light: Science \& Applications}} \textbf{\bibinfo{volume}{9}}, \bibinfo{pages}{30} (\bibinfo{year}{2020}).

\bibitem{puc2023broadband}
\bibinfo{author}{Puc, U.}, \bibinfo{author}{Yang, J.-A.}, \bibinfo{author}{Kim, D.}, \bibinfo{author}{Kwon, O.-P.} \& \bibinfo{author}{Jazbinsek, M.}
\newblock \bibinfo{title}{Broadband thz wave generation in organic benzothiazolium crystals at mhz repetition rates}.
\newblock \emph{\bibinfo{journal}{Optical Materials Express}} \textbf{\bibinfo{volume}{13}}, \bibinfo{pages}{53--66} (\bibinfo{year}{2023}).

\bibitem{gollner2021highly}
\bibinfo{author}{Gollner, C.} \emph{et~al.}
\newblock \bibinfo{title}{Highly efficient {THz} generation by optical rectification of mid-ir pulses in {DAST}}.
\newblock \emph{\bibinfo{journal}{APL Photonics}} \textbf{\bibinfo{volume}{6}} (\bibinfo{year}{2021}).

\bibitem{singh2018gapless}
\bibinfo{author}{Singh, A.}, \bibinfo{author}{Pashkin, A.}, \bibinfo{author}{Winnerl, S.}, \bibinfo{author}{Helm, M.} \& \bibinfo{author}{Schneider, H.}
\newblock \bibinfo{title}{Gapless broadband terahertz emission from a germanium photoconductive emitter}.
\newblock \emph{\bibinfo{journal}{ACS Photonics}} \textbf{\bibinfo{volume}{5}}, \bibinfo{pages}{2718--2723} (\bibinfo{year}{2018}).

\end{thebibliography}

%\begin{figure}
%\textbf{Table of Contents}\\
%\medskip
%  \includegraphics{toc-image.png}
%  \medskip
%  \caption*{}
%\end{figure}

\end{document}